\documentclass[sn-mathphys-num]{sn-jnl}

\usepackage{graphicx}%
\usepackage{multirow}%
\usepackage{amsmath,amssymb,amsfonts}%
\usepackage{amsthm}%
\usepackage{mathrsfs}%
\usepackage[title]{appendix}%
\usepackage{xcolor}%
\usepackage{textcomp}%
\usepackage{manyfoot}%
\usepackage{booktabs}%
\usepackage{algorithm}%
\usepackage{algorithmicx}%
\usepackage{algpseudocode}%
\usepackage{listings}%
\usepackage{enumitem}

\theoremstyle{thmstyleone}%
\theoremstyle{thmstyletwo}%

\theoremstyle{thmstylethree}%

\begin{document}

\title[LangYa: Revolutionizing Cross-Spatiotemporal Ocean Forecasting]{LangYa: Revolutionizing Cross-Spatiotemporal Ocean Forecasting}

\author[1,2]{\fnm{Nan} \sur{Yang}}
\author[1,2]{\fnm{Chong} \sur{Wang}}
\author[1,2]{\fnm{Meihua} \sur{Zhao}}
\author[1,2]{\fnm{Zimeng} \sur{Zhao}}
\author[1,2]{\fnm{Huiling} \sur{Zheng}}
\author[1,2]{\fnm{Bin} \sur{Zhang}}
\author[1,2]{\fnm{Jianing} \sur{Wang}}
\author*[1,2]{\fnm{Xiaofeng} \sur{Li}}\email{lixf@qdio.ac.cn}

\affil[1]{\orgdiv{Key Laboratory of Ocean Observation and Forecasting}, \orgaddress{\city{Qingdao}, \country{China}}}
\affil[2]{\orgdiv{Key Laboratory of Ocean Circulation and Waves, Institute of Oceanology}, \orgname{Chinese Academy of Sciences}, \orgaddress{\city{Qingdao}, \country{China}}}



\abstract{Ocean forecasting is crucial for both scientific research and societal benefits. Large artificial intelligence (AI)-based models have recently boosted forecasting efficiency and accuracy. However, it remains challenging to develop a comprehensive AI-driven ocean forecasting system capable of integrating cross-spatiotemporal and atmospheric forcing. This study introduces LangYa, a cross-spatiotemporal and atmospheric forcing ocean forecasting system featuring: (1) a large-language-model-based (LLM-based) time embedding to represent forecast timescale explicitly, (2) an asynchronous cross-iterative random sampling strategy to simulate atmospheric forcing impacts on ocean processes, (3) an ocean self-attention module to enhance network stability and accelerate training convergence, and (4) an adaptive loss function to capture ocean dynamics at ocean thermocline depth ranging from tens to 300 meters. Trained on 27 years of global ocean data from the Global Ocean Reanalysis and Simulation version 12 (GLORYS12). Compared to numerical and AI-based ocean forecasting systems, LangYa enables one model to make forecasts with lead times of 1 to 7 days and achieves cross-spatiotemporal forecasting at a high resolution (1/12°, daily) in ocean temperature, salinity, and current. Specifically, LangYa realizes approximately a 50\% improvement in thermocline forecast accuracy. LangYa provides ocean researchers with a powerful ocean forecasting tool and establishes a new paradigm in ocean research.}


\maketitle

\section*{Main}\label{sec1}

Ocean forecasting aims to forecast future ocean dynamic states and is crucial for understanding ocean activities~\cite{teal1984oil,schiller2020bluelink,zohdi2019harmful,breivik2013advances}. Numerical global ocean forecasting systems (GOFSs), such as the Mercator Ocean's global ocean analysis and forecasting system (PSY4) and the Real-Time Ocean Forecasting System (RTOFS), rely on geophysical fluid dynamics and partial differential equations (PDEs) to describe future ocean conditions, and such systems remain widely used worldwide~\cite{garraffo2020rtofs,francis2020high,lellouche2018mercator,tonani2015status,bauer2015quiet,smith2016sea,jean2021copernicus}.

Large artificial intelligence (AI) models are reshaping interactions between humans and technology through their massive learning parameters and powerful learning capabilities~\cite{reichstein2019deep,duben2021machine,camps2025artificial,si2024can}, e.g., ChatGPT and Sora. Such advances open the way for the development of geophysical system forecasting~\cite{wang2025advancing,wang2023deepblue,li2023artificial,li2020deep}, including large AI models for weather (e.g., FourcastNet, PanGu-Weather~\cite{ravuri2021skilful,pathak2022fourcastnet,bi2023accurate,lam2023learning,chen2023fengwu,kochkov2024neural}) and ocean (e.g., XiHe, WenHai, AI-GOMS~\cite{xiong2023ai,wang2024xihe,aouni2024glonet,cui2025forecasting}) forecasts. These large models are mainly data-driven, recasting atmosphere state variables (ASVs) and ocean state variables (OSVs), forecasting problems as autoregressive video generation, thereby avoiding much computational complexity and uncertainties inherent in physical processes. Moreover, specialized computing devices such as graphics processing units (GPUs) offer extremely fast inference speeds~\cite{ba2016layer,chen2021crossvit,touvron2022deit}. However, by investigating and practicing ocean forecasting, we summarize three key barriers in current large ocean forecasting models: (1) Future OSVs are forecasted autoregressively in a step-by-step manner, with temporal information not incorporated into the training and inference processes~\cite{xue2020coupled}. It neglects the time-dependent evolution of OSV fields and their interdependencies, making it impossible to achieve cross-spatiotemporal forecasts~\cite{wang2025advancing}. (2) Atmosphere and ocean data are concatenated as model inputs to represent their interactions, in a simple way that cannot effectively simulate the influence of atmosphere on the ocean nor integrate their underlying mechanisms into the training and inference processes~\cite{xue2020coupled}. (3) Available AI-based open-source large models rely on an ensemble of multi-sub-models for forecasting rather than using a single model to achieve multivariate forecasting across multiple oceanic variables.

Here, we present the LangYa forecasting system (see the Methods section for "LangYa"), a large AI-based model for cross-spatiotemporal, atmospheric forcing ocean forecasting that achieves high spatiotemporal resolution (1/12°, 1 day) forecasts. LangYa delivers faster and more accurate forecasts for OSVs (temperature, salinity, and currents) than both traditional numerical weather prediction (NWP) systems and XiHe (an influential and the only openly available AI model). Our contributions are fourfold: (1) We propose a large-language-model-based (LLM-based) Time Embedding Module, which leverages LLMs' prior knowledge and enables cross-spatiotemporal forecasts (up to 7 days) without iterative steps, i.e., using one model to make forecasts with lead times of 1 to 7 days. (2) We develop an asynchronous cross-iterative random sampling strategy to simulate the stochastic influence of the atmosphere on the ocean, improving the forecasting skill for variables with different rates of change, such as sea temperature and currents. (3) We introduce an ocean self-attention module based on cosine attention, ensuring more stable model training and faster convergence. With this module, a single epoch of distributed training on global OSVs from 1993 to 2019 can be completed in just 0.5 days. (4) We propose an adaptive loss function for thermocline, which significantly improves the accuracy of forecasts of ocean thermocline (tens to 300 meters), yielding about a 50\% improvement in thermocline forecast accuracy compared to other methods. Using GLORYS12 reanalysis data, we verify that LangYa can achieve stable training and convergence, and effectively yield real-time, accurate, multivariate forecasts across spatiotemporal scales, especially for the thermocline layer. The ablation study for our contribution is detailed in the supplementary file. Ablation experiments confirm that each module significantly contributes to LangYa's forecasting performance, reducing errors particularly in challenging thermocline regions and complex ocean dynamics.

\begin{figure}[!t]
    \centering
    \includegraphics[width=1.\textwidth]{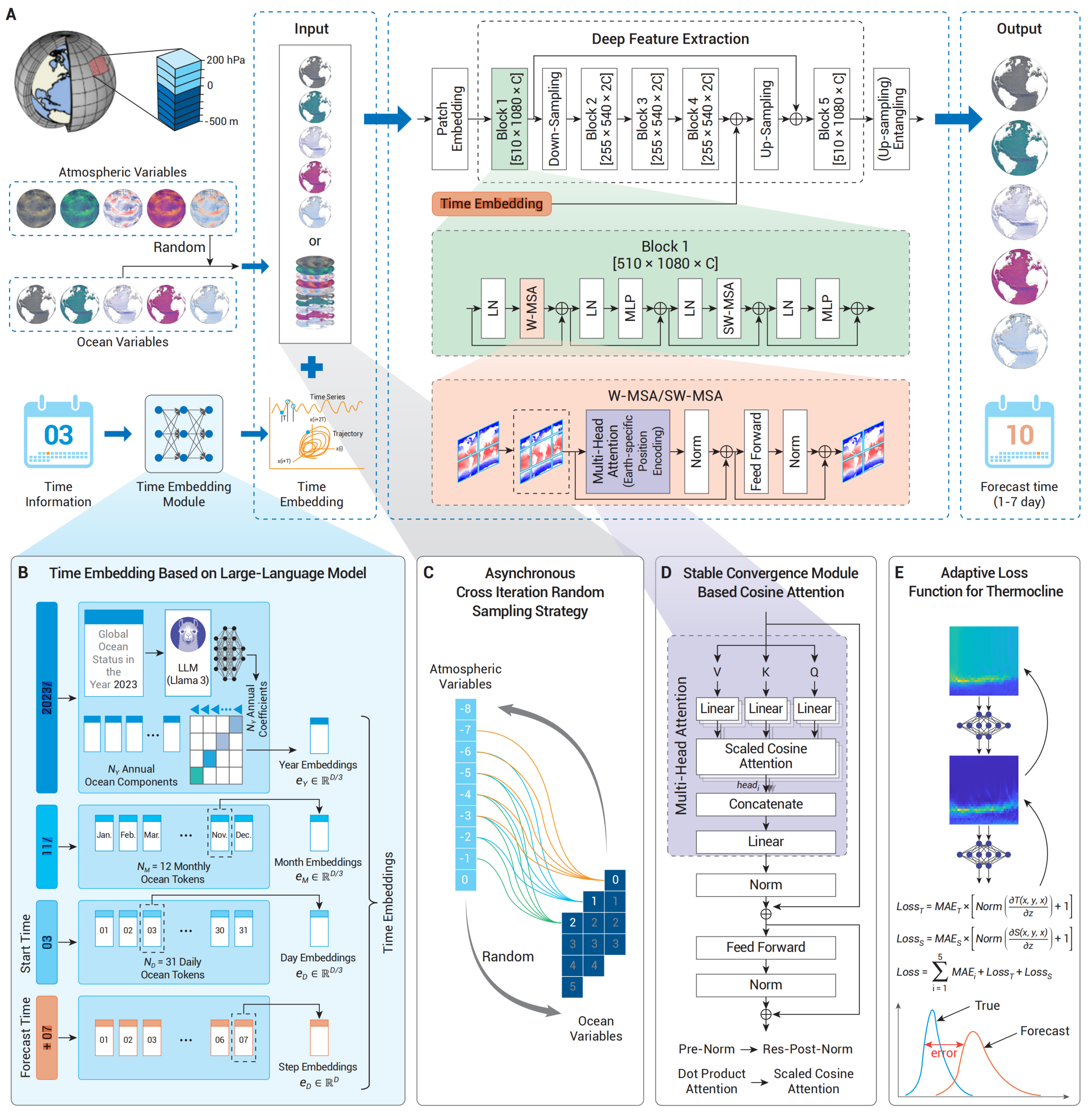}
    \caption{\textbf{Overview of the LangYa Forecasting System Architecture and Its Key Module.} (A) Overall system pipeline integrating Time Embedding and deep feature extraction, (B) LLM-based Time Embedding module, (C) Asynchronous Cross-Iterative Random Sampling Strategy, (D) Ocean self-attention module based on cosine attention, and (E) Adaptive loss function for thermocline forecasts. W-MSA means window-based multi-head self-attention module. SW-MSA means shifted window-based multi-head self-attention.}\label{fig1}
\end{figure}

\section*{Cross-SpatioTemporal and Air-Sea Coupled Ocean Forecasting System}\label{sec2}

LangYa adopts a data-driven idea, considering ocean forecasting as a video prediction task conditioned on oceanic and atmospheric forcing fields. The design contains training and testing phases by using oceanic and atmospheric reanalysis data as inputs and generating forecasts for Ocean State Variables (OSVs) — such as sea temperature, salinity, and currents — through single-pass inference over 1–7 days. The system uses 10585 samples (daily samples across the years 1993-2021) and mitigates overfitting risks by randomly shuffling sample order at the beginning of each training iteration. The training was conducted using a distributed data-parallel (DDP) strategy on a cluster of 16 NVIDIA A800 GPUs, completed in just 14 days. Fig.~\ref{fig1} illustrates the architecture (Fig.~\ref{fig1}A) and key modules (time-embedding module in Fig.~\ref{fig1}B, Asynchronous Cross-Iterative Random Sampling Strategy in Fig.~\ref{fig1}C, Ocean Self-Attention Module in Fig.~\ref{fig1}D, Adaptive Thermocline Loss Function in Fig.~\ref{fig1}E) of LangYa, a cross-spatiotemporal and atmospheric forcing ocean forecasting system.

The core architecture of the LangYa is shown in Fig.~\ref{fig1}A, which integrates multiple components. The input state variables include ASVs, which are u- and v-components of wind speed, air temperature, and relative humidity across 1000 to 200 hPa. Sea Surface State Variables (SSSVs): u- and v-components of 10-m wind speed, sea surface temperature, and mean sea level pressure. OSVs: Temperature, salinity, and currents spanning depths of 0 to -500 m. OSVs, ASVs, and SSSVs derived from the European Centre for Medium-Range Weather Forecasts (ECMWF) fifth-generation reanalysis dataset (ERA5) and GLORYS12 reanalysis datasets. The data is complemented by temporal information processed via a time-embedding module based on LLM (Fig.~\ref{fig1}B). The module extracts spatial features for individual time steps and temporal features across multiple time steps, and then integrates prior knowledge from the LLM to encode forecast durations. The unique fusion of spatiotemporal data supports highly efficient cross-spatiotemporal forecasting.

A distinctive feature of LangYa is its Asynchronous Cross-Iterative Random Sampling Strategy (Fig.~\ref{fig1}C), which adaptively links ASVs, SSSVs, and OSVs to simulate the effect of the atmosphere on the ocean. Unlike traditional methods that rely on physics-based assumptions or simple data concatenation, this strategy learns spatiotemporal relationships and effectively captures atmosphere-ocean variation patterns, significantly enhancing forecast accuracy without introducing additional physical constraints. The processed input data are passed through an encoder-decoder architecture equipped with an Ocean Self-Attention Module (Fig.~\ref{fig1}D). The module, based on cosine attention, ensures stable training and accelerates convergence by reducing gradient explosions common in high-dimensional data processing. To overcome inaccurate thermocline forecasts, LangYa incorporates an Adaptive Thermocline Loss Function (Fig.~\ref{fig1}E). The loss function is tailored to capture the steep vertical gradients and dynamic variability of the thermocline, improving forecast accuracy for this critical oceanic layer.

By integrating these components, LangYa: (1) significantly reduces temporal distribution discrepancies and simplifies the complexity of atmosphere-ocean coupling. (2) supports efficient, accurate, and stable forecasting across spatiotemporal scales. (3) yields fine performance, particularly in challenging regions like thermoclines and areas with intense atmosphere-ocean interactions. LangYa's comprehensive design represents a major leap forward in AI-driven ocean forecasting, offering unprecedented efficiency and reliability for global ocean research and operational applications.

\section*{Experimental Setting and Main Results}\label{sec3}

We trained the LangYa forecasting system using reanalysis data (GLORYS12 and ERA5~\cite{bell2021era5,saha2010research}) and evaluated it with both reanalysis (GLORYS12) and observational datasets. The observational datasets are derived from the GODAE Ocean View Inter-comparison and Validation Task Team (IV-TT) Class 4 framework~\cite{ryan2015godae}, which provides observation datasets from Argo, Jason-1, Jason-2, and Envisat CLS AVISO level 3 satellite altimeters. 27 years of GLORYS12 data (1993–2019) were used for training LangYa, while data from 2020–2021 were reserved for testing. LangYa forecasts 128 variables, including temperature, salinity, and zonal and meridional ocean currents across 32 depth layers.

\begin{figure}[!t]
    \centering
    \includegraphics[width=1.\textwidth]{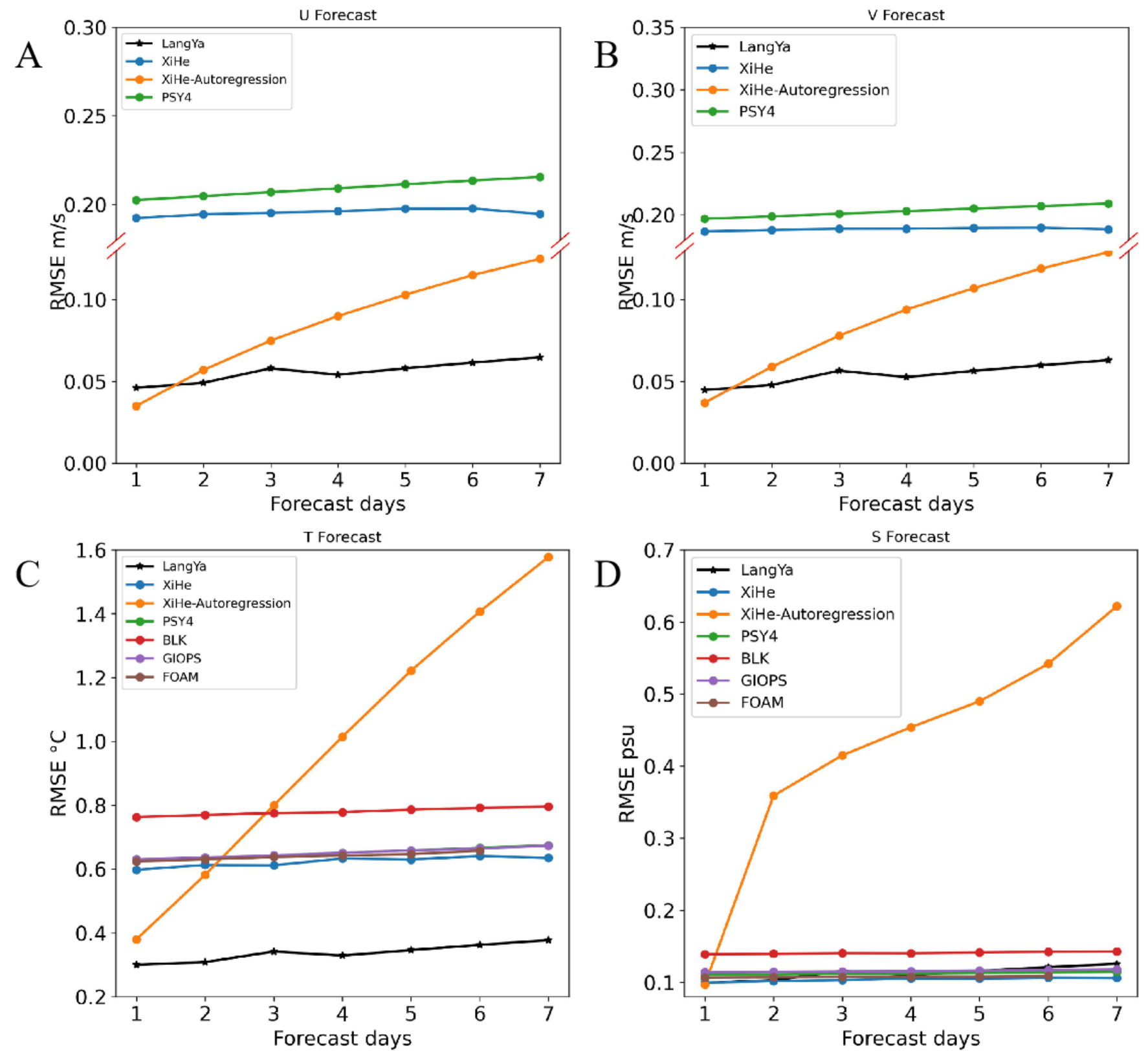}
    \caption{\textbf{Performance Comparison of LangYa, XiHe, XiHe-AR, and Numerical Models Across OSVs.}  A: RMSE of the u-component of ocean currents for 1-7 day forecasts. B: RMSE of the v-component of ocean currents for 1-7 day forecasts. C: RMSE of the ocean temperature for 1-7 day forecasts. D: RMSE of the salinity for 1-7 Day Forecasts.}\label{fig2}
\end{figure}

Under the same experimental environment configuration, compared with reanalysis and observational data, LangYa achieved lower RMSEs (Please refer to the Model Training and Metrics section in the Methods for the definition) for temperature and ocean currents (both zonal and meridional) than XiHe (the only openly available AI model), XiHe-Autoregression (XiHe-AR), PSY4, BLK, GIOPS, and FOAM numerical models. Fig.~\ref{fig2}A denotes the RMSE of the u-component of ocean currents for 1-7 day forecasts. Fig.~\ref{fig2}B is the RMSE of the v-component of ocean currents for 1-7 day forecasts. Fig.~\ref{fig2}C indicates RMSE of the ocean temperature for 1-7 day forecasts. Fig.~\ref{fig2}D denotes the RMSE of the salinity for 1-7 day forecasts. Notably, the released XiHe version consists of 20 separate models, each designed to forecast OSVs for 1–10 days in either the upper or lower ocean layers. XiHe-AR refers to an autoregressive approach that generates multiday forecasts by sequentially chaining two 1-day XiHe models, resulting in higher cumulative errors. For salinity, LangYa's forecast accuracy was comparable to XiHe and the numerical models but outperformed XiHe-AR.

In addition, LangYa achieved an inference time of 1 second for forecasts of any scale (1-7 days) on a single GPU, which is over 10,000 times faster than operational numerical models. Results from the 2020–2021 test data indicate that LangYa provided reasonable quantitative forecasts (RMSE), particularly for challenging ocean phenomena such as thermoclines. Compared to the world's leading OSVs forecasting systems, LangYa significantly reduced cumulative forecast errors and demonstrated remarkable improvements in thermocline forecasting.

\begin{figure}[!t]
    \centering
    \includegraphics[width=1.\textwidth]{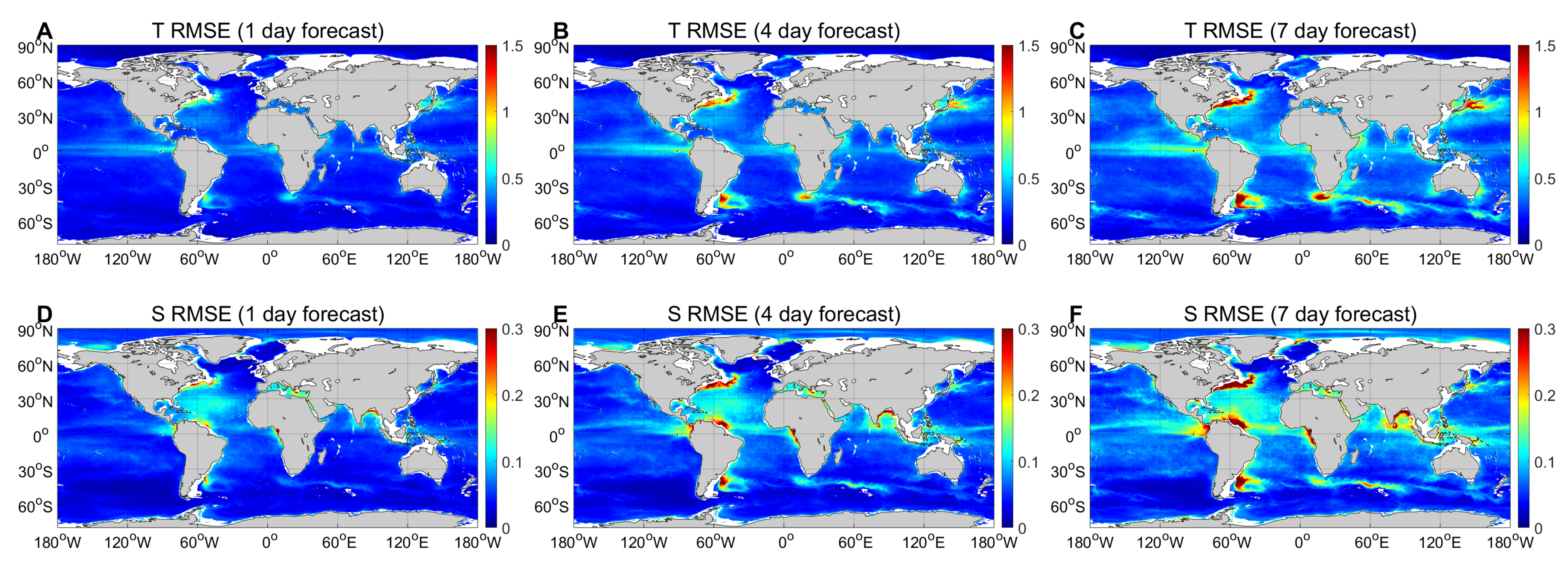}
    \caption{\textbf{Global RMSE Distribution of LangYa Forecasts.}  Temperature and Salinity at 1, 4, and 7-day forecasts. A-C: RMSE distribution of temperature at 1, 4, and 7-day forecasts. D-F: RMSE distribution of salinity at 1, 4, and 7-day forecasts.}\label{fig3}
\end{figure}

Fig.~\ref{fig3} illustrates the RMSE of LangYa's forecasts for temperature and salinity at 1, 4, and 7 days. Spatially, LangYa achieves RMSEs below 0.3°C for temperature forecasts across most ocean regions. Areas with higher RMSEs are concentrated in the western boundary currents (e.g., the Kuroshio Extension, Gulf Stream), eastern boundary currents (e.g., the Peru Current, California Current), and the Somali Current. These regions experience large temperature variations, making them challenging to forecast accurately using numerical models or XiHe. Nevertheless, LangYa demonstrates significant advantages in these areas. For salinity, LangYa achieves RMSEs below 0.05 psu across most ocean regions, particularly in open oceans. Higher RMSEs are observed near the equator, western boundary currents, and the Indonesian Throughflow region, where salinity is heavily influenced by complex processes such as evaporation, precipitation, river runoff, and water mass exchange. Despite these complexities, LangYa delivers accurate salinity forecasts in these challenging regions. Furthermore, the spatial distribution of LangYa's forecast errors for both temperature and salinity remains consistent as the forecast time increases, indicating the system's high forecast stability.

\begin{figure}[!t]
    \centering
    \includegraphics[width=1.\textwidth]{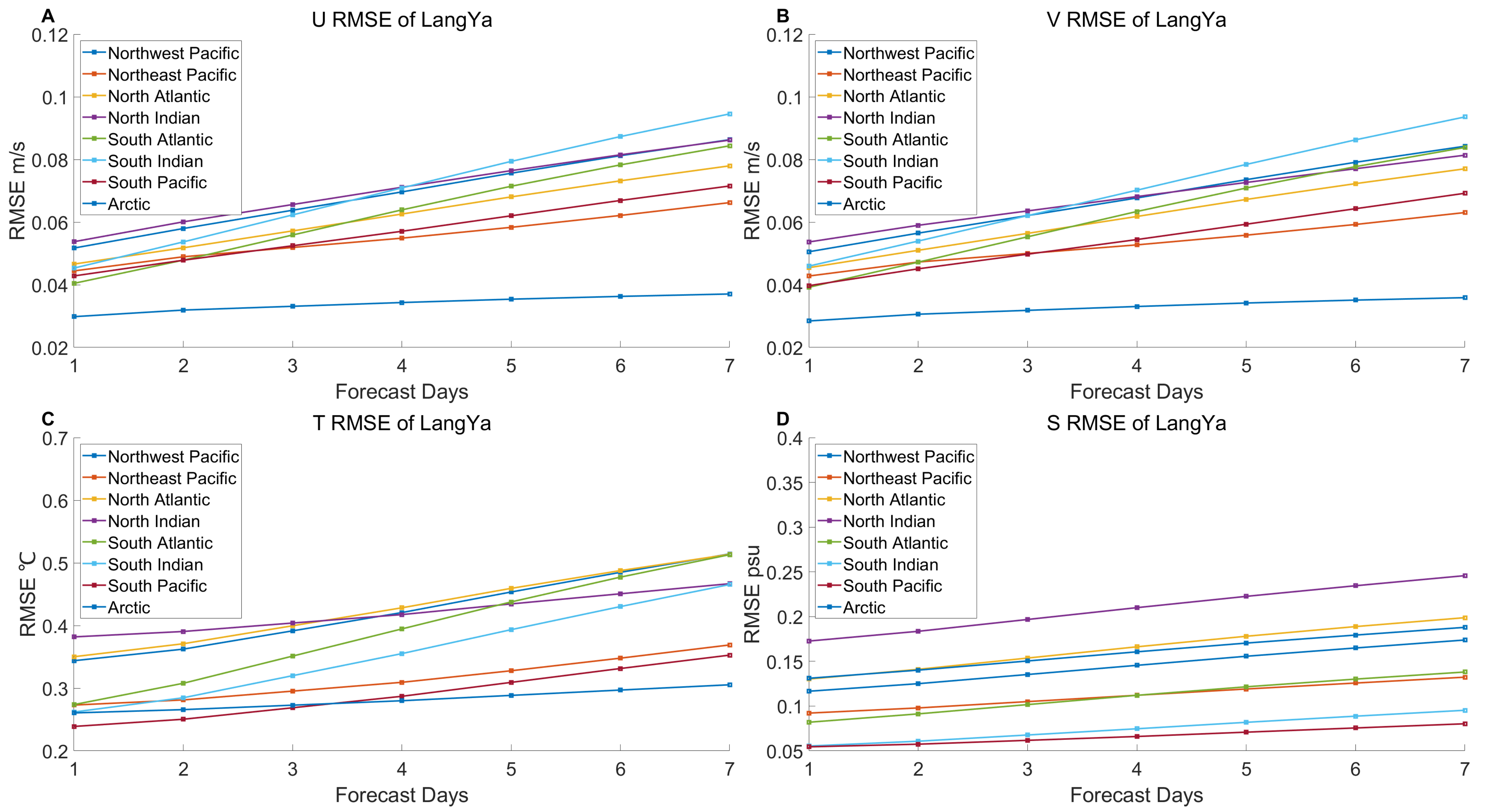}
    \caption{\textbf{RMSE of LangYa Forecasts for Temperature, Salinity, and Currents Across Eight Ocean Basins (1-7 Days).}  A-D: u-, v-component of ocean currents, temperature, and salinity forecast RMSE at different ocean basins.}\label{fig4}
\end{figure}

Fig.~\ref{fig4} presents LangYa's forecasting accuracy for four OSVs across these eight basins (Northwest Pacific, Northeast Pacific, North Atlantic, North Indian Ocean, South Atlantic, South Pacific, South Indian Ocean, and Arctic Ocean) over 1–7 days. The results show substantial variations in RMSEs among the different ocean basins for forecasting ocean variables. Specifically, the Arctic Ocean exhibited the lowest RMSEs for U, V, and T, indicating that LangYa's performance in this region is the most stable and accurate. This is due to the relatively stable hydrodynamic conditions and fewer drastic changes in the Arctic. Conversely, forecasting OSVs in the North Indian Ocean presented significant challenges, as monsoons, freshwater input, and complex oceanic dynamics influence the North Indian basin. The intricate physical processes and higher uncertainties posed difficulties for LangYa's forecasts. In contrast, LangYa demonstrated more consistent forecasting performance in the other basins, showcasing the robustness of the LangYa forecasting system.

Notably, in salinity forecasts, there were significant differences in RMSEs between the North Indian Ocean and North Atlantic (relatively high values) and the two Southern Hemisphere basins (South Pacific and South Indian Ocean, relatively low values). It can be inferred from Fig.~\ref{fig3} that the higher RMSEs in the North Atlantic are concentrated in the Gulf Stream extension region. The rapid salinity changes and spatial heterogeneity in this area are driven by freshwater input from the Gulf Stream and rivers. In the North Indian Ocean, the low accuracy of salinity forecasts is primarily due to the influence of complex monsoon systems and river runoff (e.g., the Ganges and Brahmaputra Rivers). Additionally, the presence of coastal upwelling and the Indonesian Throughflow, which intensify water mass mixing, further complicates the forecasts in this region. These areas, characterized by complex physical processes and unique oceanic conditions, highlight the need for further development and refinement of the forecasting model.

\section*{Thermocline Forecast}\label{sec4}

The thermocline, a transitional layer in the ocean characterized by a sharp temperature gradient, is a critical interface for energy exchange, water mass mixing, and biogeochemical processes. Typically located at tens to 100 m depths, the thermocline exhibits complex regional and seasonal variations due to its involvement in multiple coupled physical processes, such as ocean stratification, vertical mixing, and atmosphere-ocean interactions. Accurate forecasting of the thermocline is essential for advancing the understanding of these processes and their implications for global climate systems, yet it has long been a challenge due to the steep gradients and non-linear dynamics in this layer~\cite{chu2017exponential}. 

LangYa employs a physics-guided adaptive loss function specifically designed to address these challenges by directing the model’s attention to the complex temperature dynamics of the thermocline. This approach not only enhances forecasting accuracy but also deepens our understanding of thermocline formation, evolution, and its interactions with adjacent ocean layers.

\begin{figure}[!t]
    \centering
    \includegraphics[width=1.\textwidth]{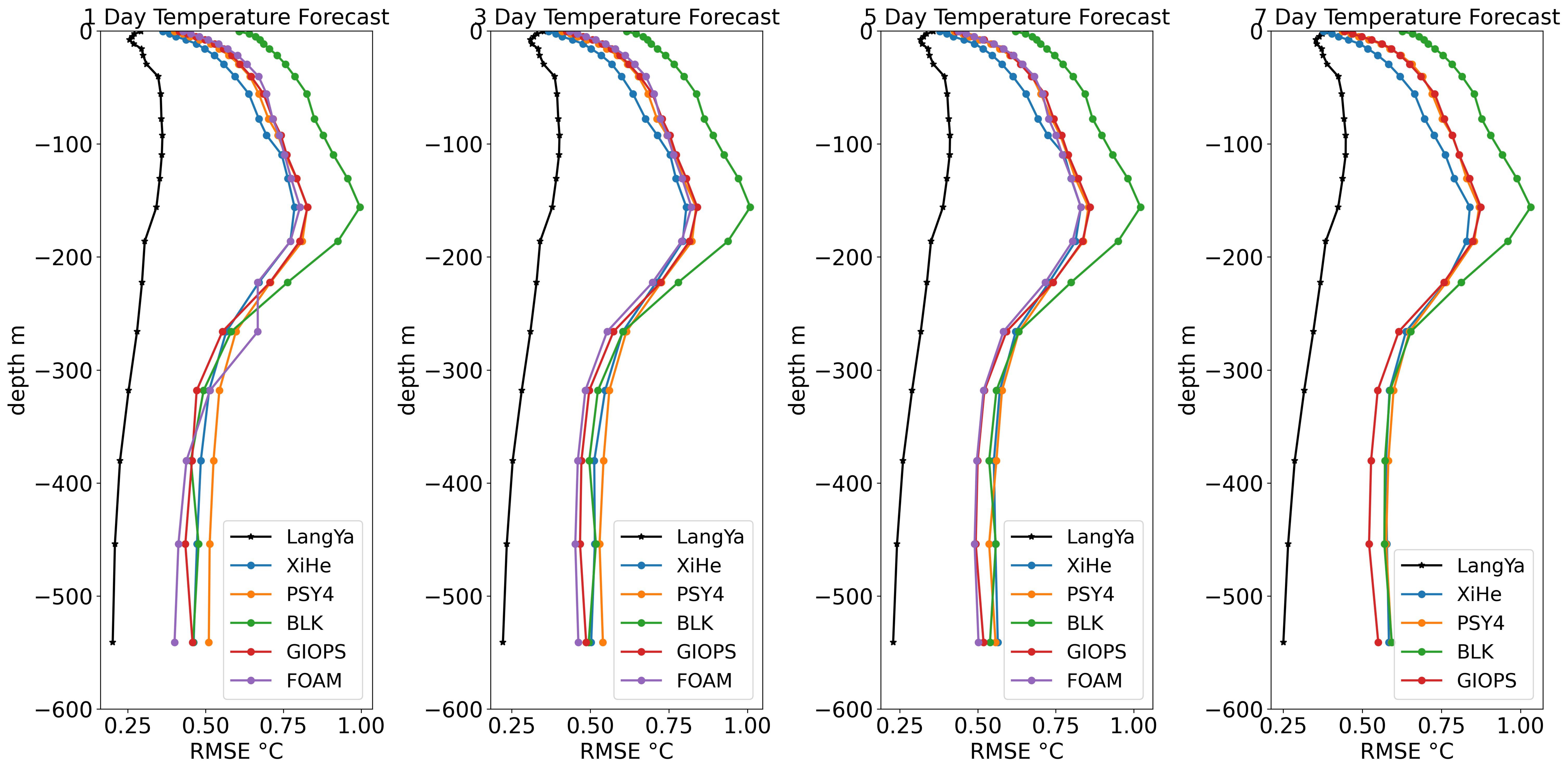}
    \caption{\textbf{Depth-Dependent RMSE Distributions of LangYa's Temperature Forecasts at 1, 3, 5, and 7 Days.} A: 1-day forecast result. B: 3-day forecast result. C: 5-day forecast result. D: 7-day forecast result.}\label{fig5}
\end{figure}

The experimental results illustrated in Fig.~\ref{fig5} demonstrate that LangYa consistently delivers superior performance in forecasting temperature profiles across all evaluated forecast horizons, outperforming both conventional numerical models and established open-source AI-based systems, such as XiHe. LangYa's advantage is especially pronounced near the thermocline, a region where traditional methods often struggle to accurately represent rapid temperature transitions. By minimizing RMSE near the thermocline, LangYa achieves unprecedented forecasting accuracy and significantly improves its ability to simulate and analyze ocean dynamics. In regions with intense atmosphere-ocean interactions and surface mixing, LangYa's performance remains robust, highlighting its ability to incorporate the stochastic and coupled nature of mixing processes. As depth increases and the model approaches the thermocline (identified by the pronounced rise in RMSE profiles), LangYa's advantage becomes even more pronounced. The experimental result suggests that the adaptive loss function not only captures the vertical gradients but also enhances the model's ability to generalize across varying oceanographic conditions.

\section*{Discussion}\label{sec5}

The LangYa forecasting system addresses critical bottlenecks in cross-spatiotemporal and atmospheric-forced ocean forecasting, as well as thermocline forecasting. By integrating Time Embedding, asynchronous cross-iterative random sampling strategy, and adaptive loss function for thermocline, LangYa demonstrates significant advantages in forecasting OSVs across temporal scales. Compared with other forecasting systems (e.g., GOFSs and XiHe), LangYa completes global OSVs inferencing within a short period and exhibits superior performance in accuracy, timeliness, and stability, offering feasibility for real-time ocean forecasting deployment.

Traditional ocean forecasting systems rely on iterative processes to estimate future OSVs, which are computationally expensive and prone to cumulative errors. LangYa overcomes the limitations by employing a Time Embedding module, enabling forecasts for up to 7 days without iterative steps. This approach enhances its cross-temporal forecasting capabilities and improves forecasting flexibility and efficiency. In terms of the effect of the atmosphere on the ocean, LangYa utilizes an asynchronous cross-iterative random sampling strategy, successfully overcoming the simplistic concatenation of atmosphere-ocean data used in existing AI-based ocean forecasting systems. The strategy effectively simulates the complex effect of the atmosphere on the ocean, significantly improving the forecasting accuracy of OSVs, such as temperature and currents. Moreover, LangYa's training and convergence stability have been substantially enhanced through the incorporation of an ocean self-attention module. The ocean self-attention module optimizes the training process, accelerates convergence, and ensures stable forecasting outcomes. 

A significant strength of LangYa is its capability to accurately forecast thermocline dynamics. The thermocline, a region in the ocean characterized by sharp temperature changes, has long been a challenging area in ocean science. By designing a loss function guided by physical knowledge, LangYa effectively captures the evolution characteristics of the thermocline, significantly improving forecasting accuracy. Experiments reveal that LangYa achieves lower RMSE in thermocline forecasts compared to existing AI-based ocean forecasting systems and numerical models, accurately reflecting the complex spatial and temporal variations of the thermocline.

LangYa's experiments across eight global ocean basins further demonstrate its superior adaptability and stability. Whether in the relatively stable Arctic Ocean or the complex monsoon- and river-runoff-influenced ocean basins, LangYa maintains high forecasting accuracy and exhibits remarkable robustness across various ocean state variables. It indicates that LangYa can yield reliable forecasts even in dynamic and uncertain ocean environments, highlighting its broad application potential.

LangYa forecasting system provides a novel tool for oceanographic and Earth science research, advancing the integration of AI technologies with ocean science. Its innovative algorithm design and robust forecasting capabilities not only lay a theoretical foundation for real-time ocean forecasting systems but also offer precise and efficient support for global climate monitoring, oceanic resource management, and disaster early warning. With continuous technological advancements and data accumulation, LangYa is poised to become a core tool in global ocean science research, contributing significantly to the development of ocean science and its practical applications.

Despite its achievements, LangYa faces several limitations in practical applications: (1) LangYa is trained and tested on reanalysis data, whereas operational forecasting systems primarily rely on numerical assimilation data. The differences between reanalysis and numerical assimilation data may reduce model performance in applications, requiring further research into cross-data-source adaptability. This is not a unique problem for LangYa but a common problem for all AI models. (2) LangYa currently supports short-term forecasts of 1–7 days and has not yet been extended to medium-term forecasts (e.g., exceeding 10-day forecasts). (3) LangYa focuses on forecasting OSVs and has not yet been extended to directly forecast ocean phenomena. Forecasts are not accurate in areas of dramatic change, so further research is needed.

\bibliography{sn-bibliography}

\section*{Methods}\label{sec6}

\subsection*{Dataset}\label{sec61}

The OSVs data used in LangYa are derived from the GLORYS12 reanalysis data. GLORYS12 was developed and implemented under the framework of the Copernicus Marine Environment Monitoring Service (CMEMS27). It has a horizontal resolution of 1/12°, 50 vertical layers, and a temporal spanning from January 1993 to December 2021. Due to its high resolution, temporal continuity, long coverage period, and excellent data quality, GLORYS12 has become the preferred dataset for large model development.

Considering the important effect of the atmosphere on the ocean in ocean dynamics, LangYa incorporates ASVs (u- and v-components of wind speed, temperature, and relative humidity at 200, 500, 850, and 1000 hPa) and SSSVs (u- and v-components of 10-m wind speed, sea surface temperature, and mean sea level pressure) in the design. The ASVs and SSSVs data are sourced from the ERA533, which has a spatial resolution of 0.25° and includes 37 vertical pressure levels.

\subsection*{Data Preprocessing}\label{sec62}

The input data for LangYa includes 149 variables, categorized as follows: (1) 5 SSSVs: sea surface temperature, sea surface height, mean sea level pressure, u-, and v-components of 10-m wind speed. (2) 128 OSVs: These are distributed across 32 vertical depth layers (0.49, 1.54, 2.65, 3.82, 5.08, 6.44, 7.93, 9.57, 11.41, 13.47, 15.81, 18.50, 21.60, 25.21, 29.44, 34.43, 40.34, 47.37, 55.76, 65.81, 77.85, 92.32, 109.73, 130.67, 155.85, 186.13, 222.48, 266.04, 318.13, 380.21, 453.94, and 541.09 m), with four variables per depth layer: ocean temperature, salinity, u-, and v- components of ocean currents. (3) 6 ASVs: These are distributed across four pressure levels (200, 500, 850, and 1000 hPa), with four variables per level: air temperature, relative humidity, u-, and v-components of wind speed.

The study focuses on a global domain with longitude spanning -180° to 180° and latitude spanning -80° to 90°. The grid dimensions for OSVs are H=4320, W=2041. For consistent resolution, the ERA5 data were interpolated to a spatial resolution of 1/12° to match the GLORYS12 reanalysis data using the nearest-neighbor interpolation method. The nearest-neighbor interpolation method was selected for its simplicity and efficiency, while preserving spatial consistency and spatiotemporal dependencies between grid points.

For each input channel, LangYa applies channel-wise Z-Score normalization to transform the data into a standard normal distribution with a mean of 0 and a standard deviation of 1.

\subsection*{LangYa Forecasting System (Fig.~\ref{fig1}A)}\label{sec63} 

LangYa was trained on 27 years of data from 1993 to 2019 and tested using data from 2020 and 2021. LangYa inputs OSVs, ASVs, and SSSVs, to forecast OSVs (channels, C=128) with a spatial resolution of 1/12°. As shown in Fig.~\ref{fig1}A, the forecasting process can be expressed as:

\begin{equation}
    \hat{\mathrm{X}}_{\mathrm{t} +\tau}=\mathcal{F}_\theta\left(\mathrm{H}\left(\mathrm{X}_{\mathrm{t}}, \mathrm{~A}_{(\mathrm{t}-10): \mathrm{t}}\right) \mid \mathrm{E}(\mathrm{t}, \tau)\right)
\end{equation}

where $\mathrm{X}_{\mathrm{t}} \in \mathbb{R}^{(D_X, H, W)}$ represents the OSVs at time t inputted to the forecasting system. It includes temperature, salinity, and ocean currents at 32 depth levels, which are combined along the channel dimension. $\mathrm{~A}_{(\mathrm{t}-10): \mathrm{t}} \in \mathbb{R}^{(D_A, H, W)}$ represents the historical ASVs over the previous 10 days, forming the contextual input for the forecasting system(The study primarily focuses on short-term forecasts of OSVs at a daily time resolution.). $\hat{\mathrm{X}}_{\mathrm{t} \tau}$ denotes the LangYa forecast for OSVs at time $\tau \in \{1,2, ..., K\}$. To standardize notation, $\mathrm{X}^*_{\mathrm{t}+\tau}$ represent the corresponding ground truth at the timestep $(\mathrm{t}+\tau)$. $\mathcal{F}_\theta(\cdot)$ denotes LangYa's ocean self-attention module. $\mathrm{H}(\cdot)$ represents LangYa's modeling of air-sea interaction mechanisms, specifically implemented via the asynchronous cross-iterative random sampling strategy. $\mathrm{E}(\mathrm{t}, \tau)$ encodes the temporal information of LangYa's initial and forecast horizons.

The specific forecasting steps of LangYa are as follows: \begin{enumerate}[label=(\arabic*)]
    \item Coupling of OSVs, ASVs, and SSSVs: LangYa integrates OSVs, ASVs, and SSSVs using the asynchronous cross-iterative random sampling strategy within the atmospheric forcing ocean module. The process generates data with dimensions 2041x4320x133. Subsequently, the Time Embedding module, based on LLMs, encodes the temporal information of the current and forecast times into a 96-dimensional feature vector. These two features are combined to form the input to LangYa.
    \item Spatiotemporal Feature Extraction: LangYa employs an ocean self-attention module to extract spatiotemporal features, capturing the complex dependencies across spatial and temporal dimensions in the input data. For OSVs, LangYa utilizes a Patch Embedding module to perform dimensionality reduction by dividing the input grid into non-overlapping patches, with each patch encoded into a latent representation of dimension C. During experiments, the patch size is set to 4×4, resulting in feature dimensions of 510×1080×C. 
    \item Deep Feature Extraction via Swin Transformer Blocks (ST Blocks): LangYa uses five ST blocks~\cite{liu2021swin} to extract deep features from the patch tokens. The ST blocks are organized hierarchically, where the number of tokens is progressively reduced through downsampling layers and then restored step by step through upsampling layers. Time Embeddings are incorporated into the ST blocks through tensor addition. 
    \item Upsampling for Forecast Generation: LangYa employs an upsampling module composed of interpolation and convolution operations to generate forecasted OSVs with dimensions 2041×4320×128.
\end{enumerate}

The technical details of the deep feature extraction architecture and the ST Block are as follows: \begin{enumerate}[label=(\arabic*)]
    \item Deep Feature Extraction Architecture: For the first ST block, the input data has dimensions of 510×1080×C. In the subsequent three layers, the spatial dimensions are reduced by half, while the number of channels is doubled, resulting in data dimensions of 255×540×2C. In the final ST block, the data dimensions are restored to 510×1080×C. During the downsampling process, 4 tokens are merged into 1, increasing the feature dimension from C to 4C, and then reduced to 2C via a linear layer. During upsampling, the reverse operation is performed, restoring the data to its original dimensions.
    \item Swin Transformer Block (ST Block): Each ST block consists of a Window-based Multi-Head Self-Attention (W-MSA) module followed by a Shifted Window-based Multi-Head Self-Attention (SW-MSA) module~\cite{liu2022swin}. Ocean data typically has extremely high spatial resolution, which imposes a significant computational burden on classical Transformers. To improve computational efficiency, the W-MSA module divides the input data into multiple small windows and applies the self-attention mechanism within each window for feature extraction. In our experiments, the window size is set to 5×12. However, window-based self-attention lacks connections between windows, which limits its modeling capacity. To introduce cross-window connections while maintaining the computational efficiency of non-overlapping windows, LangYa employs the SW-MSA module, alternating W-MSA and SW-MSA modules within the ST block. In the ST block, the attention computation is performed only within each small window. Considering that the values of OSVs are closely related to the absolute geographic location on Earth, we incorporate Earth-specific positional encoding. The attention mechanism is computed as follows:
    \begin{equation}
    \text { Attention }(\mathrm{Q}, \mathrm{~K}, \mathrm{~V})=\operatorname{SoftMax}\left(\frac{\mathrm{QK}^{\mathrm{T}}}{\|\mathrm{Q}\|_2\|\mathrm{~K}\|_2}+\mathrm{B}\right) \mathrm{V}
    \end{equation}
    where Q represents the query, K the key, V the value, and B the Earth-specific positional bias.
\end{enumerate}

\subsection*{Time Embedding Module (Fig.~\ref{fig1}B)}\label{sec64}

There are two methods for forecasting long-term time series of OSVs and ASVs: (1) autoregressive forecasting methods (e.g., Pangu-weather) and (2) building separate forecasting models for each forecast day (e.g., XiHe). In general, the first method leads to significant cumulative errors, while the second entails high training costs due to the need for multiple models. Neither method satisfies the requirement of using a single model for cross-temporal forecasting of OSVs, highlighting the urgent need to overcome this bottleneck.

Inspired by recent progress in the denoising diffusion probabilistic model (DDPM~\cite{ho2020denoising}), LangYa employs time embedding as a carrier of temporal information to encode both the initial time and the forecast intervals. It enables a single model to achieve cross-temporal forecasting of OSVs for various initial times and forecast days. The periodic patterns exhibited by OSVs over years, seasons, and months are encoded into the initial time embedding. On the other hand, the nonlinear temporal variations over different forecast lead times are correspondingly encoded into the interval time embedding. The combination of the above two items constitutes a complete time-embedded formalization:
\begin{equation}
E(t, \tau) \triangleq E\left(t_Y \oplus t_M \oplus t_D, \tau\right)=E_Y\left(t_Y\right) \oplus E_M\left(t_M\right) \oplus E_D\left(t_D\right)+E_\tau(\tau)
\end{equation}
where Y, M, and D denote the values of year, month, and day, respectively. $\tau$ represents the forecast lead times. M and Dare represented as discrete, learnable embedding vectors. Taking into account the fact that there is no clear periodic pattern in the numbering of years and that future years are not enumerable, we represent it as the linear combination of principal component embedding vectors:
\begin{equation}
    {E}_{\mathrm{Y}}\left(\mathrm{t}_{\mathrm{Y}}\right) \triangleq \sum_{\mathrm{i}=1}^{\mathrm{N}_{\mathrm{Y}}} \mathrm{w}_{\mathrm{Y}}^{(\mathrm{i})} \cdot \mathrm{e}_{\mathrm{Y}}^{(\mathrm{i})}
\end{equation}
where $\sum_{\mathrm{i}=1}^{\mathrm{N}_{\mathrm{Y}}} \mathrm{w}_{\mathrm{Y}}^{(\mathrm{i})} = 1$ are the coefficients of the principal component vectors. These coefficients are derived from LLM with prior knowledge and are mapped to the required dimensions using a multilayer perceptron (MLP): 
\begin{equation}
    \left\{\mathrm{w}_{\mathrm{Y}}^{(\mathrm{i})}\right\}^{\mathrm{N}}{ }_{\mathrm{i}=1}=\operatorname{MLP}\left[\operatorname{LLM}\left(\mathrm{t}_{\mathrm{Y}}\right)\right]
\end{equation}
In practice, the instruction-tuned Llama3.1-8B model is adopted as the LLM backbone. Interaction with the LLM is guided by the prompt: "Global Ocean Status in the Year XXX." To derive embeddings that effectively capture context rather than individual word semantics, the hidden state embeddings from the final layer of the Llama model are utilized.

\subsection*{Atmospheric Forcing Ocean Module (Asynchronous Cross-iterative Random Sampling Strategy, Fig.~\ref{fig1}C)}\label{sec65}

The variations in OSVs do not occur in isolation but are driven by the complex interactions between the ocean and the atmosphere. These coupled relationships pose significant challenges for high-precision forecasting of OSVs, particularly under limited computational resources. Many existing models either fail to simultaneously handle ASVs, SSSVs, and OSVs data or inadequately account for their relationships, relying solely on simple data concatenation for modeling. Additionally, the inconsistency in spatial resolutions between ASVs, SSSVs, and OSVs data further complicates their integration within models.

To address the challenges, we propose a sampling scheme (asynchronous cross-iterative random sampling strategy) for training to simulate the effect of the atmosphere on the ocean. The strategy simulates the random effect of the atmosphere on the ocean by randomly sampling ASVs, SSSVs, and OSVs input data during the training process. The goal is to use ASVs data to achieve an accurate forecasting of OSVs and SSSVs. The core concept of the strategy is based on the fundamental fact that the temporal scale of atmospheric changes is significantly faster than that of oceanic changes. In other words, oceanic conditions on any given day can be substantially affected by atmospheric states from preceding days. Accordingly, we define an operator $\sum_{k=t-10}^t \mathrm{H}\left(\mathrm{~A}_{\mathrm{k}}\right) * \mathrm{~A}_{\mathrm{k}}$, indicating that ASVs influence the forecast of OSVs on a given day from the previous 1 to 10 days. The process is formalized as:
\begin{equation}
    \mathrm{H}\left(\mathrm{X}_{\mathrm{t}}, \mathrm{~A}_{(\mathrm{t}-10): \mathrm{t}}\right) \triangleq \mathrm{X}_{\mathrm{t}} \oplus \mathrm{SS}_{\mathrm{t}} \oplus \sum_{k=t-10}^t \mathrm{H}\left(\mathrm{~A}_{\mathrm{k}}\right) * \mathrm{~A}_{\mathrm{k}}
\end{equation}
Here, SS refers to the SSSVs, while A denotes the ASVs. $\mathrm{H}(\cdot)$ learns to map 16-layer upper-air variables within 10 days as 10 learning weights. Each weight is a scalar and controls the extent to which atmospheric variables have mixed with ocean variables over the past 10 days. $\mathrm{A}_{\mathrm{k}}$ denotes the previous atmospheric data on the $k$'th day. $\mathrm{H}\left(\mathrm{~A}_{\mathrm{k}}\right)$ denotes the learned weight on the $k$'th day. Specifically, the mechanism of the strategy works as follows: for the forecast of oceanic variables at the current time (t), OSVs may be influenced by ASVs between time $(t-10)$ and $(t-1)$. A specific time within the range is determined through random sampling. Similarly, at the current time $(t+1)$, the influence shifts to ASVs between time $(t-9)$ and $t$. The sliding-window-based random sampling mechanism effectively captures the stochasticity and complexity of atmosphere-ocean interactions. The sampling strategy is integrated into the multi-node, multi-GPU training process and executed asynchronously. The design not only improves computational efficiency but also preserves the complex dynamic coupling relationships among the data. Therefore, it is called the Asynchronous Cross-Iterative Random Sampling Strategy.

More importantly, the approach accounts for the lead-lag coupling effects between the ocean and atmosphere, enabling the model to fully learn atmosphere-ocean coupling across different temporal scales during training: (1) During actual forecasting, the coupling days can be adaptive learning based on real-time oceanic conditions. (2) The influence of the atmosphere on the ocean is treated as a stochastic perturbation. The time scales $(t-1)$ to $(t-10)$ are coupled through learnable weights. Therefore, we can use the ensemble forecasting approach to achieve stable forecasting results, which improves the robustness of LangYa's forecasts.

\subsection*{Ocean Self-Attention Module}\label{sec66}

OSVs, SSSVs, and ASVs often have very high resolutions, which makes gradient explosion (i.e., the rapid increase of deep activation values) more common during large model training. To address this issue, we propose a stability-enhancing module based on cosine attention. Two key modifications were made: 

(1) We replaced the previous pre-normalization architecture with a post-normalization structure in the residual blocks. Thus, the output of each residual block is normalized before merging back into the main branch. The design ensures that the amplitude of the main branch does not accumulate as the network depth increases. The computation of each ST block is given by:
\begin{equation}
    \begin{aligned}
        & \hat{z}^1=\operatorname{LN}\left(\mathrm{W}-\operatorname{MSA}\left(\mathrm{z}^{\mathrm{l}-1}\right)\right)+\mathrm{z}^{\mathrm{l}-1} \\
        & \mathrm{z}^{\mathrm{l}}=\operatorname{LN}\left(\operatorname{MLP}\left(\hat{\mathrm{z}}^{\mathrm{l}}\right)\right)+\hat{\mathrm{z}}^{\mathrm{l}} \\
        & \hat{\mathrm{z}}^{\mathrm{l}+1}=\operatorname{LN}\left(\operatorname{SW}-\operatorname{MSA}\left(\mathrm{z}^{\mathrm{l}}\right)\right)+\mathrm{z}^{\mathrm{l}} \\
        & \mathrm{z}^{\mathrm{l}+1}=\operatorname{LN}\left(\operatorname{MLP}\left(\hat{\mathrm{z}}^{\mathrm{l}+1}\right)\right)+\hat{\mathrm{z}}^{1+1}
    \end{aligned}
\end{equation}
where $\mathrm{z}^{\mathrm{l}-1}$ denotes the input to the ST block, $\mathrm{z}^{\mathrm{l}+1}$ denotes the output of the block, "W-MSA"  represents the window-based multi-head self-attention computation, SW-MSA represents the shifted-window multi-head self-attention computation, and LN denotes layer normalization.

(2) Ocean forecasting requires handling multidimensional input data, such as OSVs, ASVs, and SSSVs. The dot-product attention mechanism can be affected by the scale variations of these high-dimensional inputs, leading to instability during training. To address it, we replaced the original dot-product attention with a scaled cosine attention mechanism (named ocean self-attention module), which ensures that the computation is independent of the input amplitude at each block. 

\subsection*{Adaptive Thermocline Loss Function (Fig.~\ref{fig1}E)}\label{sec67}

The thermocline is a transition layer in the ocean where temperature changes drastically with depth. Its formation mechanism is complex, making accurate thermocline forecasting a significant challenge in ocean modeling. The difficulty of forecasting the thermocline lies in the sharp gradient changes in water temperature within this region, which are often difficult for traditional models to learn and capture effectively. To address this, we developed an adaptive thermocline loss function to enhance the model's learning capability in the thermocline. The loss function is expressed as:
\begin{equation}
    \mathcal{L}_{\mathrm{T}}=\left|\left[\mathrm{X}_{\mathrm{T}}\right]_{\mathrm{t}}^*-\left[\hat{\mathrm{X}}_{\mathrm{T}}\right]_{\mathrm{t}}\right| \times\left[\operatorname{Norm}\left(\frac{\partial\left[\mathrm{X}_{\mathrm{T}}\right]_{\mathrm{t}}^*(\mathrm{x}, \mathrm{y}, \mathrm{z})}{\partial \mathrm{z}}\right)+1\right]
\end{equation}
where $[\mathrm{X}_{\mathrm{T}}]$ represents the grid for temperature, and $0<x<W, 0<y<H, 0<z<C_T$ represents the grid coordinates for longitude, latitude, and depth. Norm denotes the min-max normalization, scaling the data to a range of $[0,1]$. $\left[\mathrm{X}_{\mathrm{T}}\right]_{\mathrm{t}}^*, \left[\hat{\mathrm{X}}_{\mathrm{T}}\right]$ represent the corresponding ground truth from the reanalysis and LangYa's forecast values. The loss function enables the model to focus more on regions with significant vertical gradients, particularly the temperature variations near the thermocline, thereby significantly improving its ability to learn the complex evolution of the thermocline.

Based on the definition above, the overall training loss for LangYa is defined as:
\begin{equation}
    \mathcal{L}=\left|\mathrm{X}_{\mathrm{t}}^*-\hat{\mathrm{X}}_{\mathrm{t}}\right|+\lambda \mathcal{L}_{\mathrm{T}}
\end{equation}
where $\lambda=1.0$ is the weight for the thermocline loss.

\subsection*{Model Training and Metrics}\label{sec68}

LangYa was trained to forecast times ranging from 1 to 7 days. The asynchronous cross-iterative random sampling strategy considers atmospheric influences on the ocean for a period of 1 to 10 days. LangYa was trained for 100 epochs using the Adam optimizer, with the training executed over 14 days on a cluster of 16 NVIDIA Tesla-A800 GPUs. The batch size was set to 16, with an initial learning rate set at 0.001 and gradually decayed to zero using a cosine annealing strategy. To mitigate overfitting, the training dataset (spanning 1993–2019) was randomly shuffled within each epoch, whereas no random shuffling or sampling was applied during testing.

The study follows a recent work (XiHe) that adopted RMSE to measure the accuracy of forecasted OSVs compared to ground truth. To provide more objective experimental comparisons, MAE and PSNR are introduced as additional evaluation metrics. RMSE and MAE evaluate the global forecasting accuracy of variables from a grid-level perspective, while PSNR captures the quality of the forecast data relative to the ground truth. The formula for calculating RMSE is defined as follows:
\begin{equation}
    \begin{gathered}
    \operatorname{MAE}\left(\mathrm{X}_{\mathrm{t}}^*, \widehat{\mathrm{X}}_{\mathrm{t}}\right)=\frac{1}{\mathrm{C} \cdot \mathrm{H} \cdot \mathrm{~W}} \sum_{\mathrm{z}, \mathrm{x}, \mathrm{y}}^{\mathrm{C}, \mathrm{H}, \mathrm{~W}}\left|\mathrm{X}_{\mathrm{t}}^*(\mathrm{z}, \mathrm{x}, \mathrm{y})-\widehat{\mathrm{X}}_{\mathrm{t}}(\mathrm{z}, \mathrm{x}, \mathrm{y})\right| \\
    \operatorname{MSE}\left(\mathrm{X}_{\mathrm{t}}^*, \widehat{\mathrm{X}}_{\mathrm{t}}\right)=\frac{1}{\mathrm{C} \cdot \mathrm{H} \cdot \mathrm{~W}} \sum_{\mathrm{z}, \mathrm{x}, \mathrm{y}}^{\mathrm{C}, \mathrm{H}, \mathrm{~W}}\left\|\mathrm{X}_{\mathrm{t}}^*(\mathrm{z}, \mathrm{x}, \mathrm{y})-\widehat{\mathrm{X}}_{\mathrm{t}}(\mathrm{z}, \mathrm{x}, \mathrm{y})\right\|_2^2 \\
    \operatorname{RMSE}\left(\mathrm{X}_{\mathrm{t}}^*, \widehat{\mathrm{X}}_{\mathrm{t}}\right)=\sqrt{\operatorname{MSE}\left(\mathrm{X}_{\mathrm{t}}^*, \widehat{\mathrm{X}}_{\mathrm{t}}\right)} \\
    \operatorname{PSNR}\left(\mathrm{X}_{\mathrm{t}}^*, \widehat{\mathrm{X}}_{\mathrm{t}}\right)=10 \cdot \log _{10} \frac{\max \left(\mathrm{X}_{\mathrm{t}}^*\right)^2}{\operatorname{MSE}\left(\mathrm{X}_{\mathrm{t}}^*, \widehat{\mathrm{X}}_{\mathrm{t}}\right)}
    \end{gathered}
\end{equation}
The results calculated using the above metrics are averaged across all time steps (testing set time range) and horizontal grid points to produce the mean RMSE, MAE, and PSNR values for the forecasted OSVs at a lead time. These metrics also remain robust in local regions, allowing for evaluation of RMSE, MAE, and PSNR for specific basins, such as Northwest Pacific, Northeast Pacific, etc.

\subsection*{Comparative Methods}\label{sec69}

Four numerical systems (PSY4, BLK, GIOPS, and FOAM) and one AI system (XiHe) that can be used for 1/125 numerical global thermohaline and current OSV forecasts are included in our comparison experiment.

The methods used to evaluate the model include the AI-based global OSVs forecasting system XiHe. XiHe achieves a high spatial resolution of approximately 1/12° and can complete a 10-day forecast within 0.36 seconds, thousands of times faster than traditional numerical Global Ocean Forecast Systems (GOFS). The evaluation also includes datasets from the IV-TT Class 4 framewor~\cite{ryan2015godae}, which provides observational datasets from Argo, Jason-1, Jason-2, and Envisat CLS AVISO level 3 satellite altimeters. 

Additionally, the IV-TT Class 4 framework includes forecast results from numerical models (PSY4, BLK, GIOPS, and FOAM) at observational points for comparative validation. PSY4: Developed by Mercator Océan in France, this model features a high spatial resolution of approximately 1/12° and a forecast duration of 10 days. BLK offers a spatial resolution of 1/10° with a forecast duration of 7 days. GIOPS: Developed by Environment and Climate Change Canada (ECCC), this model has a spatial resolution of approximately 1/4° and a forecast duration of 10 days. FOAM: Developed by the Institute of Atmospheric Physics (IAP) of the Chinese Academy of Sciences, it features a spatial resolution of 0.5° with a forecast duration of 6 days.

\section*{Data availability}\label{sec7}

For training and testing LangYa, we downloaded the GLORYS12 dataset from \url{https://data.marine.copernicus.eu/product/GLOBAL_MULTIYEAR_PHY_001_030/services} and ERA5 dataset from \url{https://cds.climate.copernicus.eu/}. For comparison with other methods, we downloaded the IV-TT Class 4 framework from \url{https://thredds.nci.org.au/thredds/catalog/rr6/intercomparison_files/catalog.html}. All these data are publicly available for research purposes. Source data will be provided in this paper.

\section*{Code availability}\label{sec8}

The code base of LangYa was established on PyTorch, a Pythonbased library for deep learning. The details of LangYa, including network architectures, modules, optimization tricks and hyperparameters, are available in the paper and the pseudocode. We will release the trained models, inference code and the pseudocode of details to the public at a GitHub repository: \url{https://github.com/iocaswolfteam/LangYa_v1_0}. The trained models allow the researchers to explore LangYa's ability on either GLORYS12 and ERA5 initial fields.

\section*{Acknowledgments}\label{sec9}

This work was supported by the National Natural Science Foundation of China (42306214), Taishan Scholars Program, Shandong Province Postdoctoral Innovative Talents Support Program (SDBX2022026), China Postdoctoral Science Foundation(2023M733533), Special Research Assistant Project of the Chinese Academy of Sciences in 2022.

\section*{Author Contributions}\label{sec10}
N.Y., C.W., Z.Z., and M.Z. conceived the study. N.Y., C.W., Z.Z., M.Z., and H.Z. developed the methodology. N.Y., X.L., and J.W. conducted the investigation. Visualization was carried out by N.Y., C.W., Z.Z., and M.Z. X.L., J.W., and B.Z. supervised the project. N.Y., C.W., Z.Z., and M.Z. wrote the original draft, and X.L. reviewed and edited the manuscript.

\section*{Competing Interests}\label{sec11}
The authors declare no competing interests.

\end{document}